\documentclass{jpsj-suppl}
\usepackage{txfonts} 

\title{Application of matrix product states to
the Hubbard model in one spatial dimension}

\author{Yukihiro \textsc{Shimizu}$^{1}$, Koji \textsc{Matsuura}$^{1}$ and Hikaru \textsc{Yahagi}$^{1}$}

\inst{$^{1}$Department of Applied Physics, Tohoku University, Sendai 980-8579, Japan}

\email{yshmz@snow.apph.tohoku.ac.jp}

\recdate{July 12, 2013}

\abst{We investigate the application of matrix product states to the
Hubbard model in one spatial dimension with both of open and periodic
boundary conditions.
We develop the variatinal method
that the optimization of the variational parameters
is carried out locally and sequentially
in the framework of matrix product operators~(MPO) by including the sign,
due to the anti-commutation relation of fermion operators,
in the matrix element of MPO.
The numerical accuracy of the ground state energy is examined.
}

\kword{matrix product states, Hubbard model, matrix product operators}

\begin{document}
\maketitle

\section{Introduction}

The variational method using so-called matrix product states~(MPS)
as a trial wave function has been developed to understand physics in low
dimensional correlated quantum systems\cite{mps1, mps2}.
The method has allowed a deeper
understanding of details and problems of the density-matrix
renormalization group method, which is believed as one of the
most powerful tools for the correlated quantum systems in one spatial
dimension.  The method of the MPS was originally developed to simulate
the quantum spin systems. There are many investigations by using the
MPS for the spin systems, however, there are not many applications to the
fermionic systems.
The negative sign due to the anti-commutation relation of the
fermion operator has been treated in the framework of the multi-scale
entanglement renormalization~(MERA)\cite{mera1, mera2} or the projected entangled pair
states~(PEPS)\cite{peps1, peps2}.
In this paper we explain a method that the sign can be exactly
treated in the framework of the matrix product operators~(MPO).
Our method is relatively simple,
and it has
applicability to the model with long range hopping, suggesting that the
treatment of sign in the MPO may be efficient even in the model in two
spatial dimension.
We will examine the numerical accuracy of the ground state energy.

\section{Matrix Product States and Matrix Product Operators}
\subsection{Open Boundary Condition}

We consider the Hubbard model of $L$ sites in one spatial dimension
with the open boundary condition.
In order to find a ground state of the model,
we employ the variational approach.
The trial function, so-called MPS,
is assumed as,
\begin{equation}
\left|\psi \rangle \right.
= \sum_{\alpha_1, \alpha_2, \cdots, \alpha_L}
A^{\alpha_1}A^{\alpha_2}\cdots A^{\alpha_L}
\left|\alpha_1, \alpha_2, \cdots, \alpha_L \rangle \right.,
\label{mps1}
\end{equation}
where $\alpha_\ell=0, \uparrow, \downarrow$ or $(\uparrow\downarrow)$ shows
the $d$-dimensional local states ($d=4$ for the Hubbard model is called
physical dimension)
at site, $\ell$.
The matrices of $\{A^{\alpha_\ell}\}$
are the variational parameters,
that the dimensions of them are
$(1\times d), (d \times {\rm min}(d^2,D)),
\cdots,
({\rm min}(d^{L/2-1},D) \times {\rm min}(d^{L/2},D))$,
$
({\rm min}(d^{L/2},D) \times {\rm min}(d^{L/2-1},D)),
\cdots,
({\rm min}(d^2,D)\times d),(d\times 1)$,
going from $\ell=1$ to $L$.
The quantity of $D$ is the bond dimension,
which controls the width of variational space.

In order to optimize $A^{\alpha_\ell}$ locally and sequentially,
we decompose the Hamiltonian into the MPO as,
\begin{align}
{\cal H}_{\rm OBC}&=-t\sum_{\ell=1, \sigma}^{L-1}
\left(c_{\ell \sigma}^\dagger c_{\ell+1 \sigma}
+ {\rm h.c.}
\right)
+ U\sum_{\ell=1}^L n_{\ell \uparrow}n_{\ell \downarrow}
=\prod_{\ell=1}^LW^{[\ell]},\\ 
W^{[1]} &=
\begin{pmatrix}
d_1 & tc_{1\uparrow}& tc_{1 \downarrow} & -tc_{1 \uparrow}^\dagger & -tc_{1 \downarrow}^\dagger & I
\end{pmatrix},\\ 
W^{[2 \le \ell \le L-1]} &=
\begin{pmatrix}
I & 0 & 0 & 0 & 0 & 0 \\
c_{\ell \uparrow}^\dagger & 0 & 0 & 0 & 0 & 0 \\
c_{\ell \downarrow}^\dagger & 0 & 0 & 0 & 0 & 0 \\
c_{\ell \uparrow} & 0 & 0 & 0 & 0 & 0 \\
c_{\ell \downarrow} & 0 & 0 & 0 & 0 & 0 \\
d_\ell & tc_{\ell \uparrow} & tc_{\ell \downarrow} & -tc_{\ell \uparrow}^\dagger & -tc_{\ell \downarrow}^\dagger & I
\end{pmatrix}, \qquad 
W^{[L]} =
\begin{pmatrix}
I \\
c_{L \uparrow}^\dagger \\
c_{L \downarrow}^\dagger \\
c_{L \uparrow}\\
c_{L \downarrow}\\
d_L
\end{pmatrix},
\end{align}
where $I$ is a unit matrix of $d \times d$,
and $d_\ell=U n_{\ell \uparrow}n_{\ell \downarrow}$.
The matrix elements of ${\cal H}_{\rm OBC}$ are given as,
\begin{align}
&\langle \alpha_1, \cdots, \alpha_L | {\cal H}_{\rm OBC} | \alpha_1', \cdots, \alpha_L' \rangle
= \sum_{b_1,\cdots,b_{L-1}}\prod_{\ell =1}^L
W^{\alpha_\ell \alpha_\ell'}_{b_{\ell-1} b_\ell}, \\
& W^{\alpha_\ell \alpha_\ell'}_{b_{\ell-1} b_\ell}
\equiv (-1)^{sign}
\langle 0 |
\left(c_{\ell \downarrow}\right)^{n_{\ell\downarrow}(\alpha_\ell)}
\left(c_{\ell \uparrow}\right)^{n_{\ell\uparrow}(\alpha_\ell)}
\left(
W^{[\ell]}
\right)_{b_{\ell-1} b_\ell}
\left(c_{\ell \uparrow}^\dagger\right)^{n_{\ell\uparrow}(\alpha_\ell')}
\left(c_{\ell \downarrow}^\dagger\right)^{n_{\ell\downarrow}(\alpha_\ell')}
|0\rangle,
\end{align}
where $(W^{[\ell]})_{b_{\ell-1} b_\ell}$
is the $(b_{\ell-1}, b_\ell)$-th matrix element of $W^{[\ell]}$.
In order to satisfy the anti-commutation relation of fermion operators,
the sign, $(-1)^{sign}$, should be given as,
\begin{align}
(-1)^{sign} &=
\left\{ \begin{array}{ll}
-1 & \text{if $W^{\alpha_1 \alpha_1'}_{1 2}$, $W^{\alpha_1 \alpha_1'}_{1 3}$, $W^{\alpha_1 \alpha_1'}_{1 4}$ and $W^{\alpha_1 \alpha_1'}_{1 5}$
with $\alpha_1' = \uparrow$ or $\downarrow$} \\
+1 & \text{otherwise}
\end{array}
\right.,
\end{align}
for $\ell=1$,
and
\begin{align}
(-1)^{sign} &=
\left\{ \begin{array}{ll}
-1 & \text{if $W^{\alpha_\ell \alpha_\ell'}_{6 2}$, $W^{\alpha_\ell \alpha_\ell'}_{6 3}$, $W^{\alpha_\ell \alpha_1'}_{6 4}$ and $W^{\alpha_\ell \alpha_\ell'}_{6 5}$
with $\alpha_\ell' = \uparrow$ or $\downarrow$} \\
+1 & \text{otherwise}
\end{array}
\right.,
\end{align}
for $2 \le \ell \le L-1$,
and $(-1)^{sign} =+1$ for $\ell=L$, respectively.

When we optimize $A^{\alpha_\ell}$ sequentially
from $\ell=1$ to $L$, and return, and so on, 
($\ell=1, 2, \cdots, L-1, L, L-1, \cdots, 2, 1, 2, \cdots$),
the problem that the minimize of
$I(\{A^{\alpha_\ell}\})
=\langle \psi | {\cal H} |
\psi\rangle
/ \langle \psi | \psi \rangle
$
is changed efficiently into a solution of the simple eigen problem
of the matrix.
We call the path of optimization as a {\it round trip}.
The optimization process of $A^{\alpha_\ell}$ is similar to the optimal
algorithm for spin models that is reviewed in section 6 in
Ref.~\cite{mps2}.
There is no numerical instability at the iterative calculation.
The detail of the optimization will be published elsewhere.

\subsection{Periodic Boundary Condition}

We consider two types of the MPS for the periodic boundary condition:
One, $|\psi^{\rm (I)} \rangle$, is identical with the type of MPS for the open boundary condition,
as shown in equation (\ref{mps1}).
Since the matrix product is obtained from the singular value
decomposition~(SVD) of the eigen vector,
if the Hamiltonian could be diagonalized, 
it is still natural that the form of $|\psi^{\rm (I)} \rangle$ is
assumed as the trial function for the periodic boundary condition.
The other, $|\psi^{\rm (II)} \rangle$, is given as,
\begin{equation}
|\psi^{({\rm II})} \rangle
= \sum_{\alpha_1, \alpha_2, \cdots, \alpha_L}
{\rm Tr}(A^{\alpha_1}A^{\alpha_2}\cdots A^{\alpha_L})
\left|\alpha_1, \alpha_2, \cdots, \alpha_L \rangle \right.,
\end{equation}
where the dimension of all matrices, $A^{\alpha_\ell}$,
is assumed as $D\times D$.
The site independent of dimension of variational parameters
may be practical advantage.
However, a cost to make the matrix elements,
$\langle \psi^{({\rm II})} | H_{\rm PBC} |\psi^{({\rm II})} \rangle$,
increases as $dD^5$ compared to $dD^3$ by assuming type-I MPS, $|\psi^{({\rm I})} \rangle$.

The MPO for the periodic boundary condition is given as
\begin{align}
{\cal H}_{\rm PBC}&={\cal H}_{\rm OBC}
-t\sum_{\sigma}
\left(c_{L \sigma}^\dagger c_{1 \sigma}
+ {\rm h.c.}
\right)
=\prod_{\ell=1}^LW^{[\ell]} + \prod_{\ell=1}^LW'^{[\ell]},\\ 
W'^{[1]} &=
\begin{pmatrix}
tc_{1\uparrow}& tc_{1 \downarrow} & -tc_{1 \uparrow}^\dagger & -tc_{1 \downarrow}^\dagger
\end{pmatrix},\\ 
W'^{[2\le\ell\le L-1]} &=
\begin{pmatrix}
I & 0 & 0 & 0 \\
0 & I & 0 & 0 \\
0 & 0 & I & 0 \\
0 & 0 & 0 & I
\end{pmatrix}, \qquad 
W'^{[L]} =
\begin{pmatrix}
c_{L \uparrow}^\dagger \\
c_{L \downarrow}^\dagger \\
c_{L \uparrow}\\
c_{L \downarrow}
\end{pmatrix},
\end{align}
where $W'^{[1]}$ and $W'^{[L]}$ denote the transfer term between $\ell=1$ and $L$.
The matrices, $W'^{[2\le\ell\le L-1]}$, are introduced to treat the anti-commutation relation of fermion operators.
The matrix elements of ${\cal H}_{\rm PBC}$ are given as,
\begin{align}
&\langle \alpha_1, \cdots, \alpha_L | {\cal H}_{\rm PBC} | \alpha_1', \cdots, \alpha_L' \rangle
= \sum_{b_1,\cdots,b_{L-1}}\prod_{\ell =1}^L
W^{\alpha_\ell \alpha_\ell'}_{b_{\ell-1} b_\ell}
+\sum_{b_1,\cdots,b_{L-1}}\prod_{\ell =1}^L
W'^{\alpha_\ell \alpha_\ell'}_{b_{\ell-1} b_\ell},\\
&W'^{\alpha_\ell \alpha_\ell'}_{b_{\ell-1} b_\ell}
\equiv (-1)^{sign}
\langle 0 |
\left(c_{\ell \downarrow}\right)^{n_{\ell\downarrow}(\alpha_\ell)}
\left(c_{\ell \uparrow}\right)^{n_{\ell\uparrow}(\alpha_\ell)}
\left(
W'^{[\ell]}
\right)_{b_{\ell-1} b_\ell}
\left(c_{\ell \uparrow}^\dagger\right)^{n_{\ell\uparrow}(\alpha_\ell')}
\left(c_{\ell \downarrow}^\dagger\right)^{n_{\ell\downarrow}(\alpha_\ell')}
|0\rangle.
\end{align}
In order to satisfy the anti-commutation relation of fermion operators,
the sign, $(-1)^{sign}$, should be given as,
\begin{align}
(-1)^{sign} =
\left\{ \begin{array}{ll}
-1 & \text{if ($1 \le \ell \le L-1$) and ($\alpha_\ell' = \uparrow$ or $\downarrow$)} \\
+1 & \text{otherwise}
\end{array}
\right..
\end{align}
We note that the treatment of the sign in
$W'^{\alpha_\ell \alpha_\ell'}_{b_{\ell-1} b_\ell}$
is one of the examples to deal with the long range hopping in a Hamiltonian.
In the case of the next nearest neighbor or the third nearest neighbor
hopping, an extension of $W$, instead of the formula by $W'$,
may be made easier.

At a naive notion,
the optimization of $A^{\alpha_\ell}$ in an around path
is the most appropriate method for the periodic boundary condition.
We will examine the numerical accuracy whether the ground state energy
depends on optimization paths,
one is the {\it round trip}, and the other
is the {\it around}.

\section{Numerical Results}

\begin{figure}[t]
 \begin{minipage}{0.5\hsize}
  \begin{center}
\includegraphics*[width=0.85\linewidth,clip]{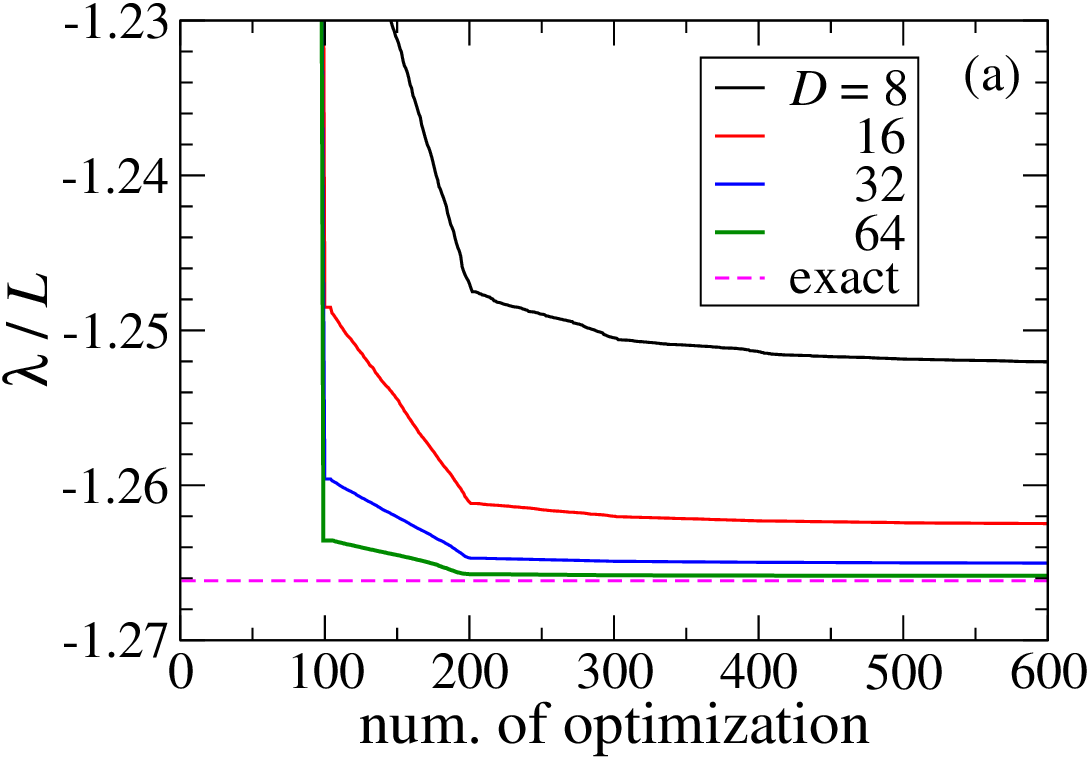}
  \end{center}
 \end{minipage}
 \begin{minipage}{0.5\hsize}
  \begin{center}
\includegraphics*[width=0.85\linewidth,clip]{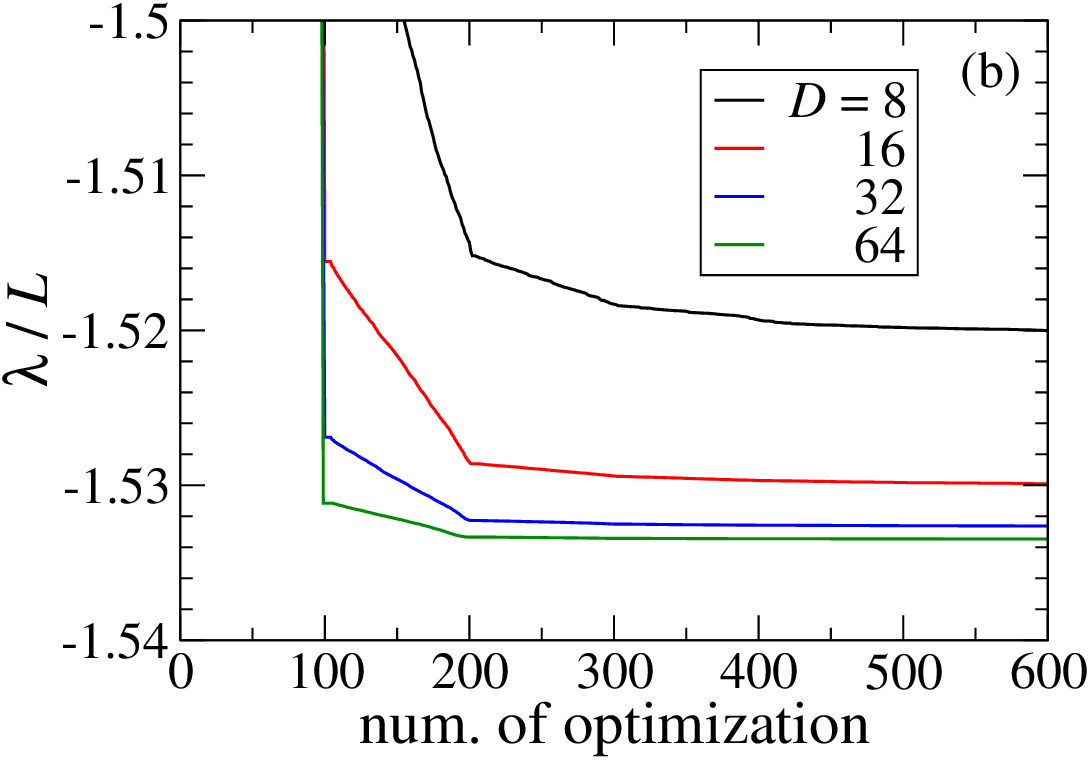}
  \end{center}
 \end{minipage}
  \caption{Optimization process of the ground state energy of the
 Hubbard model with open boundary condition.
The system size is $L=102$. The parameters of the model are
$t=1$, $U=0$ for left figure and $U=1$ for right figure.
}
  \label{fig1}
\end{figure}

As the first test we calculate the ground state energy of
non-interacting case, $U=0$, of $L=102$ sites
for the open boundary condition.
The optimization processes for various bond dimensions
are shown in Fig.~\ref{fig1}~(a).
The energy almost converges after two times round trip optimization. 
The relative error of the ground state energy
in the case of $D=64$ is less than $3 \times 10^{-4}$.
In Fig.~\ref{fig1}~(b)
the optimization process of interacting case, $U=1$, is shown.
The convergence behavior is very similar
between non-interacting case and interacting one.
We may expect similar numerical accuracy in the interacting model.
It is consistent that the entanglement
(which is measured as the spectrum of singular values of the eigen state
for the model that the Hamiltonian of small system can be exactly
diagonalized)
becomes weak with increasing $U$.

\begin{figure}[t]
 \begin{minipage}{0.5\hsize}
  \begin{center}
\includegraphics*[width=0.85\linewidth,clip]{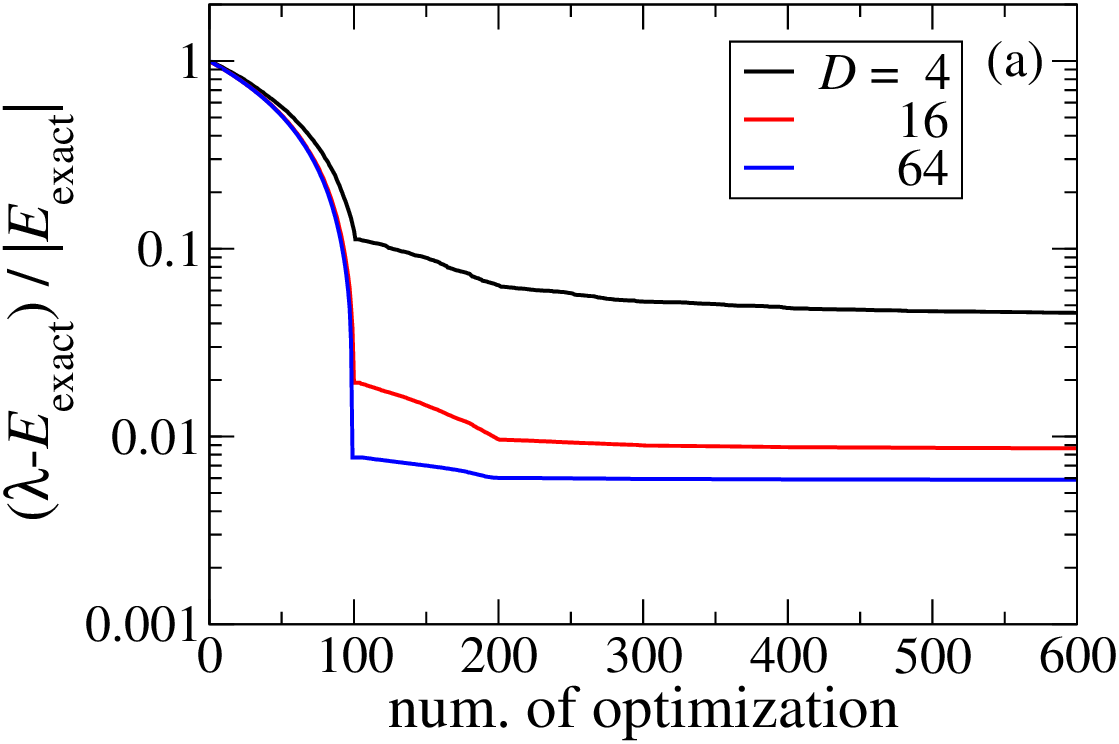}
  \end{center}
 \end{minipage}
 \begin{minipage}{0.5\hsize}
  \begin{center}
\includegraphics*[width=0.85\linewidth,clip]{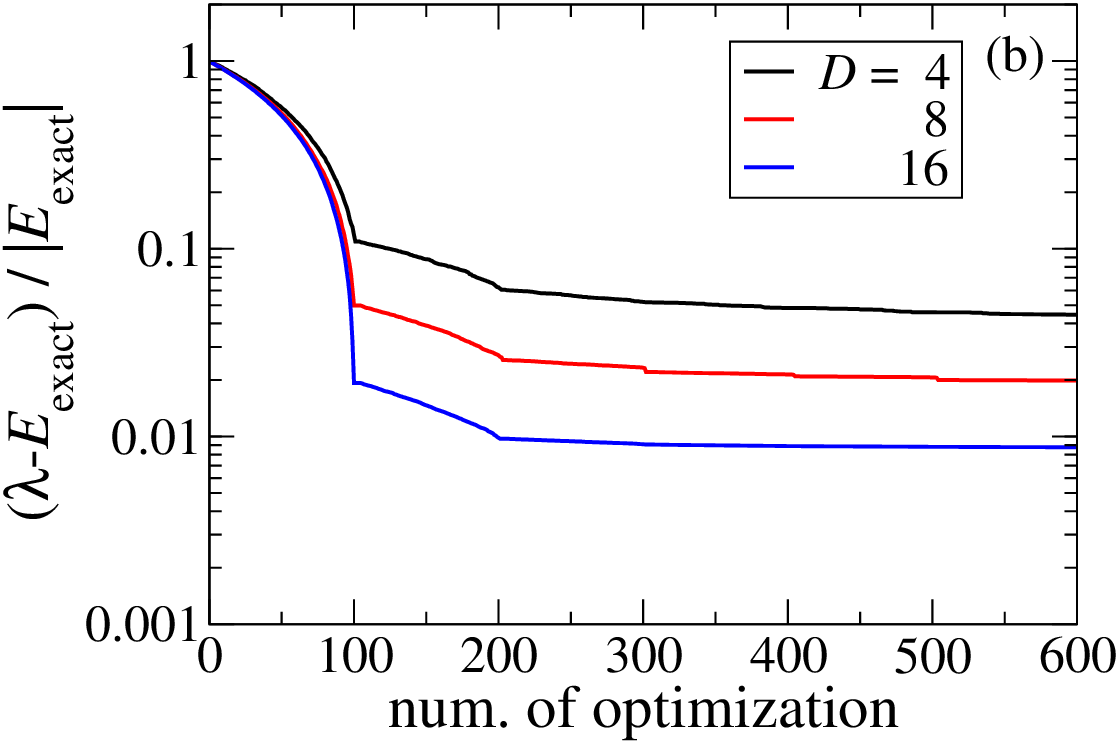}
  \end{center}
 \end{minipage}
  \caption{Relative error of the ground state energy
of non-interacting model with periodic boundary condition.
The system size is $L=102$.
The type-I trial function, $|\psi^{\rm (I)} \rangle$,
is assumed for the left figure.
The type-II trial function, $|\psi^{\rm (II)} \rangle$,
is assumed for the right figure.
}
  \label{fig2}
\end{figure}

In the case of periodic boundary condition the convergence of the ground
state energy varies in the same manner as open boundary condition
except lager $D$ is required
to reach high accuracy.
As shown in  Fig.~\ref{fig2}~(a)
the relative error of the ground state energy,
which is calculated
by employing type-I trial function, $|\psi^{\rm (I)} \rangle$,
and the {\it round trip} optimization,
is $6 \times 10^{-3}$.
If we choose the {\it around} optimization,
the relative error does not change.
When we employ the type-II trial function, $|\psi^{\rm (II)} \rangle$,
as shown in Fig.~\ref{fig2}~(b),
the numerical accuracy of the ground state energy
gets little improvement.
The numerical accuracy of the ground state properties
will be discussed in another article.
We note that
there is no numerical instability in the {\it round trip} optimization
by using type-I trial function.
In the case of type-II trial function,
there is a little instability,
but it does not much of a problem as shown in Fig.~\ref{fig2}~(b),
due to the solution of the generalized eigen problem.

\section{Conclusions}

We developed the variational method by using the MPS for the Hubbard model
with both of open and periodic boundary conditions.
The negative sign due to the anti-commutation relation of the fermion operators
can be treated in the framework of MPO.
Therefore, the variational parameters at each site
can be optimized locally.
In the case of open boundary condition
the numerical accuracy becomes well with increasing the bond dimension
of the matrix, $A^{\alpha_\ell}$.
If the fourfold bond dimension could be used,
the relative error for the non-interacting model is reduced to
one-tenth.
In the case of periodic boundary condition
the numerical accuracy does not depend on two types of trial
function and optimization paths.
In terms of the numerical cost
the best way is that the {\it round trip} optimization
with assuming trial function of $|\psi^{\rm (I)}\rangle$.
The ground state properties, such as the momentum distribution
and the spin-spin correlation function,
will be published in another article.

\end{document}